\documentclass[a4paper,11pt,openany]{article}
\usepackage{graphicx}
\usepackage[applemac]{inputenc}
\usepackage{t1enc}
\usepackage{lscape}
\usepackage{amsmath}
\usepackage{amssymb}
\usepackage{cite}
\usepackage{epsfig}
\usepackage{ulem}

\usepackage{xcolor}

\usepackage[colorlinks=true,linkcolor=blue]{hyperref}

\begin{document}

\begin{center}
\Large{\bf Validity of path thermodynamic description of

reactive systems:  Microscopic simulations}
\end{center}

\begin{center}
\large{F. Baras$^{\, a}$, Alejandro L. Garcia$^{\, b}$, and M. Malek Mansour$^{\, c}$}
\end{center}

\begin{center}
\small{ (a) Laboratoire Interdisciplinaire Carnot de Bourgogne, $\quad$\\
 UMR 6303 CNRS-Universit\'e Bourgogne Franche-Comt\'e,\\
 9 Avenue A. Savary, BP 47 870,\\
 F-21078 Dijon Cedex, France

(b) Dept. Physics and Astronomy,\\
 San Jose State University, \\
San Jose, California, 95192 USA

 (c) Universit\'e Libre de Bruxelles CP 231, Campus Plaine,\\
B-1050 Brussels, Belgium}
\end{center}

\begin{abstract}

Traditional stochastic modeling of reactive systems limits the domain of applicability of the associated path thermodynamics to systems involving a single elementary reaction at the origin of each observed change in composition.  An alternative stochastic modeling has recently been proposed to overcome this limitation. These two ways of modeling reactive systems are in principle incompatible.  The question thus arises about choosing the appropriate type of modeling to be used in practical situations. In the absence of sufficiently accurate experimental results, one way to address this issue is through the microscopic simulation of reactive fluids, usually based on hard-sphere dynamics in the Boltzmann limit. In this paper, we show that results obtained through such simulations unambiguously confirm the predictions of traditional stochastic modeling, invalidating a recently proposed alternative.

\end{abstract}

\section{Introduction}

Thermodynamic description of non-equilibrium systems is based on the concept of entropy production, the correct evaluation of which necessarily requires full knowledge of each elementary process therein \cite{Kondepudi:2008,Garcia:2022}. This prerequisite plays an important role in the corresponding statistical formulation, commonly modeled by means of an appropriate Markovian stochastic process.  In previous papers, we proved that the thermodynamic formulation of homogeneous reactive systems based on their sample path, widely known as ``path thermodynamics'', will lead to erroneous results whenever the system involves more than one elementary reaction leading to the same change in composition \cite{Malek:2020,Malek:2017}.  Similar observations were reported in \cite{DeDecker:2015} and \cite{Seleznev:2015} (see also \cite{Kurchan:1998}).

This result can also be appreciated from a physical perspective.  Let us consider a perfectly homogeneous (no local fluctuations) isothermal reactive system, such as can be produced experimentally in a  "continuously stirred tank reactor" (CSTR).  The state of such a system is entirely characterized by its composition, which is indeed the only pertinent quantity accessible to experimental investigations.  Suppose now that we have at our disposal an ideal experimental device allowing us to measure the precise number of each chemical species over an arbitrarily long interval of time.  Using this device, we may gain access to the exact state trajectory of the system, traditionally referred to as its "sample path" within the framework of stochastic processes. The question is whether such a state trajectory encompasses sufficient information to allow us to determine the thermodynamic properties of that system. The key issue here is obviously the fact that entropy production in reactive systems is the sum of the entropy production that is associated with each individual elementary reaction (cf. Section 9.5 in \cite{Kondepudi:2008}). The answer to this question will thus be positive if and only if each of the measured changes in composition can be attributed to a specific elementary reaction.  Such an attribution becomes impossible if the reactive system includes more than one elementary reaction leading to the same change in composition.  No matter what  method we use, the path thermodynamic properties of such a reactive system can never be determined from its state trajectory.

Despite the undeniability of this result, from both the mathematical and the physical point of view, its validity was recently contested by Gaspard in a Comment article \cite{Gaspard:2021}.  Yet this very issue had already been raised by that same author over a decade ago \cite{Andrieux:2007}.  In short, extending the work of Lebowitz and Spohn \cite{Lebowitz:1999}, Gaspard initiated the path thermodynamic formulation of reactive systems in 2004 \cite{Gaspard:2004,Andrieux:2004}.  The theory is sound but, like all scientific theories, its domain of applicability has limits; specific situations will exist outside the scope of the theory.  Indeed, three years later that theory was found to encounter certain inconsistencies when applied to the Schnakenberg graph formulation of "current" fluctuations in reactive systems involving more than one elementary reaction leading to the same change in composition \cite{Andrieux:2007}.  A remedy was proposed in the specific case of "current" fluctuations but, as we have shown in \cite{Malek:2020}, this remedy is inapplicable to traditional modeling of reactive systems based on jump Markov processes.

In short, a pure jump process, $\chi(t)$, is entirely determined by the concept of "transition rates", $W(X \, | \, X')$, defined by Kolmogorov as \cite{Oksendal:2003}
\begin{equation}
\label{Intro1}
P(X, t + \Delta t \, | \, X', \, t)  =  W(X \, | \,  X') \, \Delta t \, + \, {\rm o}(\Delta t) \, , \,\, \forall \, X \, \ne \, X'
\end{equation}
The function $P(X, t + \Delta t \, | \, X', t)$ represents the conditional probability to have $\chi(t + \Delta t) = X$, given that $\chi(t) = X'$.  The description of jump Markov processes in terms of their sample path is based on this fundamental Kolmogorov equality \cite{Ethier:2009}.  Consider now a reactive system that involves  $n$ elementary reactions ${\cal R}_1 \cdots {\cal R}_n$  leading the same change in composition $X \rightarrow X'$, and denote by $W_\rho(X \, | \, X')$ the transition rate associated with the reaction $\mathcal{R}_\rho$.  It was claimed in \cite{Andrieux:2007,Gaspard:2021} that we can write the sample path of $\chi(t)$ in terms of any individual transition rate $W_\rho(X \, | \, X')$.  This claim, however, contradicts the fundamental property of jump processes.

Recall that the probability associated with a random event is unique.  This property results from the basic definition of the concept of probability, first stated by Pascal, and refined over the years by various mathematicians to Kolmogorov's axiomatic formulation \cite{VanKampen:1983,Gardiner:2009}.  A transition $X' \rightarrow X$ is clearly a random event.  The probability that this event occurs in the time interval $[t, \, t + \Delta t]$  is precisely $P(X,  t + \Delta t \, | \, X',  t)$.   Accordingly, this probability cannot assume several values simultaneously, that is, for any given $X, X', t$, and $\Delta t$, the conditional probability distribution $P(X,  t + \Delta t \, | \, X',  t)$ is unique.   The Kolmogorov equality (\ref{Intro1}) then implies that this is also the case for the transition rate $W(X \, | \,  X')$.  In particular, if a reactive system involves several elementary reactions leading to the same change in composition, then the resulting transition rate is necessarily the sum of the transition rates associated with each of them, i.e.,  $W(X \, | \, X') = \sum_{\rho}W_\rho(X \, | \, X')$.  Consequently, expressing the sample path of $\chi(t)$ in terms of {\it individual} transition rates $W_\rho(X \, | \, X')$ does not apply to jump stochastic processes.

Given this limitation,  Gaspard proposed in his Comment a brand new type of stochastic modeling of reactive systems that extends the domain of validity of the associated path thermodynamics to the controversial situation of reactive systems involving more than one elementary reaction leading to the same change in composition \cite{Gaspard:2021}. The resulting stochastic process proves to be quite different from that associated with the traditional stochastic modeling of reactive systems.  They don't even share the same state space.  The state space of the stochastic process associated with the traditional modeling of an $n$ component isothermal homogeneous (CSTR) reactive system is simply $\mathbb{Z}^n$, in one-to-one correspondence with quantities that can actually be measured in laboratory experiments ($\mathbb{Z}$ represents the set of non-negative integers).  First proposed by McQuarrie in 1967 \cite{McQuarrie:1967}, traditional stochastic modeling was then refined over the years by Van Kampen \cite{VanKampen:1983}, Haken \cite{Haken:1983}, Nicolis and Prigogine \cite{Nicolis:1977}, Kurtz \cite{Kurtz:1976}, Gardiner \cite{Gardiner:2009}, and many others.  Everything we currently know about the statistical properties of non-equilibrium reactive systems was established in this way,  including the stochastic formulation of path thermodynamics by Seifert \cite{Seifert:2005},  Lebowitz and Spohn \cite{Lebowitz:1999}, and Gaspard \cite{Gaspard:2004,Andrieux:2004}.

The state space of the stochastic process associated with the new type of modeling proposed in \cite{Gaspard:2021} is different.  In addition to the number of particles of chemically active molecules, it also includes an extra variable designed to select the precise elementary reaction at the origin of an observed change in composition.  As clearly stated by the author,  the main consequence of this modeling is that it allows us to define several Markov processes associated with a given reactive system (Section III, in \cite{Gaspard:2021}).  Among them, there will undoubtedly be found a process guaranteeing the validity of the associated path thermodynamics in controversial situations.

One may argue that in this new formulation the model has been changed simply to shoehorn it to fit the theory. As they have different state spaces, it is impossible to use the framework of one of these stochastic processes to confirm or deny the validity of the other. Conversely, it is impossible to discuss the validity of the proposed new modeling within the framework of the traditional stochastic modeling.

Nevertheless, the main purpose of the two approaches is to provide a theoretical description of a "real-world" system: a system with properties that can be observed in laboratory experiments. In this respect, we know that reactive processes result from local interactions between chemically active molecules (reactive collisions). If the reactive system is well stirred (CSTR), it will be impossible to determine experimentally which reaction led to an observed change in composition, unless that reaction is unique. In other words, determining the precise state trajectory of a well-stirred reactive system, as required by the new modeling \cite{Gaspard:2021}, is beyond the reach of real-life laboratory experiments.

On the other hand, the new stochastic modeling of reactive systems, however strange, may nevertheless represent reality. The only way to investigate this possibility is through laboratory experiments. Unfortunately, the accuracy of the available experimental data are insufficient to address this issue. The alternative option is to perform microscopic simulations of reactive fluids, usually based on Newtonian hard sphere dynamics. Introduced in the mid-seventies \cite{Portnow:1975,Ortoleva:1976}, this technique provides quite useful information on the relevance and accuracy of theoretical developments in non-equilibrium reactive systems \cite{Boissonade:1979,Baras:1989,Lemarchand:1999, Hansen:2006,Baras:2004} (see \cite{Baras:1997} for a review).

In Section~2, we consider the microscopic simulation of Schl\"ogl-like reactive systems \cite{Schlogl:1971}, often used to illustrate some peculiar aspects of path thermodynamic properties of reactive systems \cite{Gaspard:2004,Andrieux:2004}. The results obtained are in perfect agreement with the traditional stochastic modeling of reactive systems, thus calling into question the alternative modeling proposed by Gaspard. Conclusions and perspectives are presented in Section~3; algorithmic details of the simulations are in an appendix.

\section{Microscopic simulation}

Let us consider the reactive system
\begin{equation}
\label{Bd1}
A \,\, + \,\, X \,\,\, \mathop{\rightleftharpoons}^{k_1}_{k_{- 1}} \, \,\, 2 \, X \quad  \quad B \,\, + \,\, C \,\,\,   \mathop{\rightleftharpoons}^{k_2}_{k_{- 2}} \, \,\, B \, + \, X
\end{equation}
where the mole fractions of the reactants $A$, $B$ and $C$ are supposed to remain constant. Following the traditional practice,  we shall use the same symbols (here $A, B, C, X$) to denote the number of particles of the corresponding chemical species.  Upon setting $k_{- 1} = k_{1}$, $k_{- 2} = k_{2}$, $C = A / 2$ and $A / B = k_{2} / k_{1} = 5 / 6$, the number of $X$ particles at the stationary state reads $X_s = A$.  This choice of parameters guaranties that the system operates under non-equilibrium conditions. For instance, we can check that in dilute (ideal) systems, the {\it thermodynamic entropy production}, $\sigma_s$, is strictly positive at the stationary state
\begin{equation}
\label{Bd2}
\sigma_s \, = \, \frac{6}{5} \, N \, k_B \, k_2 \, a^2 \,  \ln(2) \,\, > \,\, 0
\end{equation}
where $k_B$ is the Boltzmann constant, $a$ the mole fraction of $A$, and $N$ the total number of particles present in the system, including solvent or other non-reactive particles (extensivity parameter) \cite{Kondepudi:2008}.

The model (\ref{Bd1}) belongs to the class of one variable reactive systems where the reactions lead either to the change in composition $X \rightarrow X+1$  (forward) or $X \rightarrow X-1$ (backward).  A well known example is the Schl\"ogl model \cite{Schlogl:1971}.  As such, the associated state trajectory (sample path) contains no information allowing one to distinguish these reactions from each other.  We showed that this property implies necessarily that the state trajectory is time-reversible at the stationary regime, so that  the resulting {\it path entropy production} is zero \cite{Malek:2020,Malek:2017}.  It is worthwhile to recall that the time-reversibility of Schl\"ogl type of reactive systems at the stationary regime was first established by Graham and Haken in 1971 for diffusion processes \cite{Graham:1971}.  It was then generalized by several authors in order to include jump processes as well (see for example \cite{VanKampen:1983}).  An exhaustive demonstration is given in Section 6.3 of Gardiner's textbook \cite{Gardiner:2009}.

The validity of this result was contested by Gaspard \cite{Gaspard:2021}, who proposed  a new type of stochastic modeling of reactive systems where the resulting path entropy production satisfies, in average, the macroscopic thermodynamic prediction (\ref{Bd2}).  In the absence of sufficiently accurate experimental results, an alternative way to clarify this issue is through the microscopic simulation of the reactive system (\ref{Bd1}).  Measuring the number of $X$ particles over a sufficiently long time interval allows us to estimate the joint probability distribution $P(X, t \, ; X', t + \tau)$, for a given $\tau > 0$, and compare it with the reverse probability distribution $P(X', t \, ; X, t + \tau)$.  The main challenge with this approach is that obtaining statistically reliable results requires an extremely  large number of data points.

The simplest way to deal with this problem is to resort to the simulation of the reactive Boltzmann equation for which there exists a well-established algorithm introduced half a century ago by G.A.~Bird \cite{Bird:1976}.  Not only is this technique much faster than traditional hard spheres molecular dynamics, but it can also be adapted easily to simulate a perfectly homogeneous systems while conserving the main characteristics of microscopic dynamics.  Even though the latter procedure is a well-established and widely known algorithm \cite{Garcia:2000}, for the sake of completeness we present in Appendix A a short review of its basic features.

For the microscopic simulation we consider an assembly of $N = 5,000$ hard spheres of diameter $d$ confined in a box of volume $V$ with a number density of $3 \times 10^{ - 3}$ particles per $d^3$. This choice guaranties that the system is well within the range of validity of Boltzmann equation (see ref. \cite{Bird:1976} for more details). The other parameters are set as follows: $A = 1000$ particles (thus $B = 1200$ and $C = 500$ particles) and $k_2 = 0.75 \times \nu$, where $\nu$ is the collision frequency.  In other words, $75 \%$ of collisions between $A$ and $B$ or $B$ and $X$ particles (second reaction) are assumed to be reactive  (recall that $k_1 = 6/5 \, k_2 = 0.9 \, \nu$).   Note that, by definition, the number of solvent particles is changing over time since their role is to maintain constant the number of $A$, $B$ and $C$ particles (cf. Appendix A for details). We verified that the solvent mole fraction remains always above $20 \%$ during the simulation.

In addition, we considered a slightly different reactive system:
\begin{equation}
\label{Bd3}
A \,\, + \,\, A \,\,\,  \mathop{\rightleftharpoons}^{k_1}_{k_{- 1}} \, \,\, X \,\, + \,\, X  \quad ; \quad B \,\, + \,\, C \,\,\,  \mathop{\rightleftharpoons}^{k_2}_{k_{- 2}} \, \,\, B \, + \, X
\end{equation}
Unlike the reactive system (\ref{Bd1}), here each elementary reaction leads to a different change in composition so that the corresponding state trajectory is expected to be time-irreversible in a non-equilibrium stationary regime.  The comparison of the statistical properties of these two systems proves to be quite helpful for the correct interpretation of the results.  Setting $k_1 = k_2 = k_{-1} = k_{- 2}$, $B = 5 \, A /3$ and $C = A /3$, the macroscopic number of $X$ particles at the stationary state takes the same value as in system (\ref{Bd1}), that is $X_s = 1000$.  Furthermore,
the stationary state entropy production $\sigma_s$ is strictly positive.  For instance, in ideal systems,  we find
\begin{equation}
\label{Bd4}
\sigma_s \, = \, \frac{5}{9} \, N \, k_B \, k_1 \, a^2 \,  \ln(9/2) \,\, > \,\, 0
\end{equation}
so that, as before, the system operates under non-equilibrium conditions.

For the microscopic simulation of (\ref{Bd3}), the total number of hard spheres is set to $N = 7000$ with the same number density as in system (\ref{Bd1}), i.e.,  $3 \times 10^{ - 3}$ per $d^3$.  The other parameters are as follows: $A = 1500$ particles (thus $B = 2500$ and $C = 500$ particles) and $k_1 = 0.9 \times \nu$ (recall that $k_1 = k_2 = k_{-1} = k_{-2}$).

As stated above, our purpose in performing these simulations is to estimate the probability distributions $P(X, t \, ; X', t + \tau)$ and $P(X', t \, ; X, t + \tau)$.  To this end, successive values of $X$, separated by a relatively small time interval $\tau$, were recorded for statistical analysis.  For each reactive system, (\ref{Bd1}) and (\ref{Bd3}), we collected a total of $10^9$ sample points (SP), after the stationary regime has been reached (at about $10^4$ SP).

The results are presented in Figure 1.  This figure shows $P(X, t \,\, ; \, X_{ref}, t + \tau)$ and $P(X_{ref}, t \,\, ; \, X, t + \tau)$ as a function of $X$, for both systems (\ref{Bd1}) and (\ref{Bd3}), where $X_{ref} = 1000$ (i.e., macroscopic stationary state value in both systems) and $\tau = 10$ mean reactive collision time (MRCT).  The statistical error, estimated from 10 successive runs of  $10^8$ SP, does not exceed $1 \%$ (less than the size of the marker symbols in Figure 1).  While the sample path associated with model (\ref{Bd3}) is distinctly time-irreversible, as expected, this is not the case of that associated with model (\ref{Bd1}) where $P(X, t \,\, ; \, X_{ref}, t + \tau) \approx P(X_{ref}, t \,\, ; \, X, t + \tau)$.  This observation is further confirmed in Figure 2 where the ratio $P(X, t \, ; \, X_{ref}, t + \tau) / P(X_{ref}, t \, ; \, X, t + \tau)$  is depicted for both models.

The time-reversible symmetry of a sample path is the key signature of systems at thermodynamic equilibrium.  But the system (\ref{Bd1}) operates under strict non-equilibrium conditions (cf. eq. (\ref{Bd2})).  Consequently, the thermodynamic properties of this reactive system based on its sample paths will necessarily lead to a zero entropy production at the stationary-state, in obvious contradiction with assertions presented in the Comment \cite{Gaspard:2021}.  The situation is different for the case of the second model (\ref{Bd3}) since in this reactive system each elementary reaction leads to a different change in composition.   The microscopic simulation results thus confirm perfectly the validity of our results \cite{Malek:2020,Malek:2017}.

\clearpage

\begin{figure}[h!]
\begin{center}
\epsfclipon
\epsfig{file=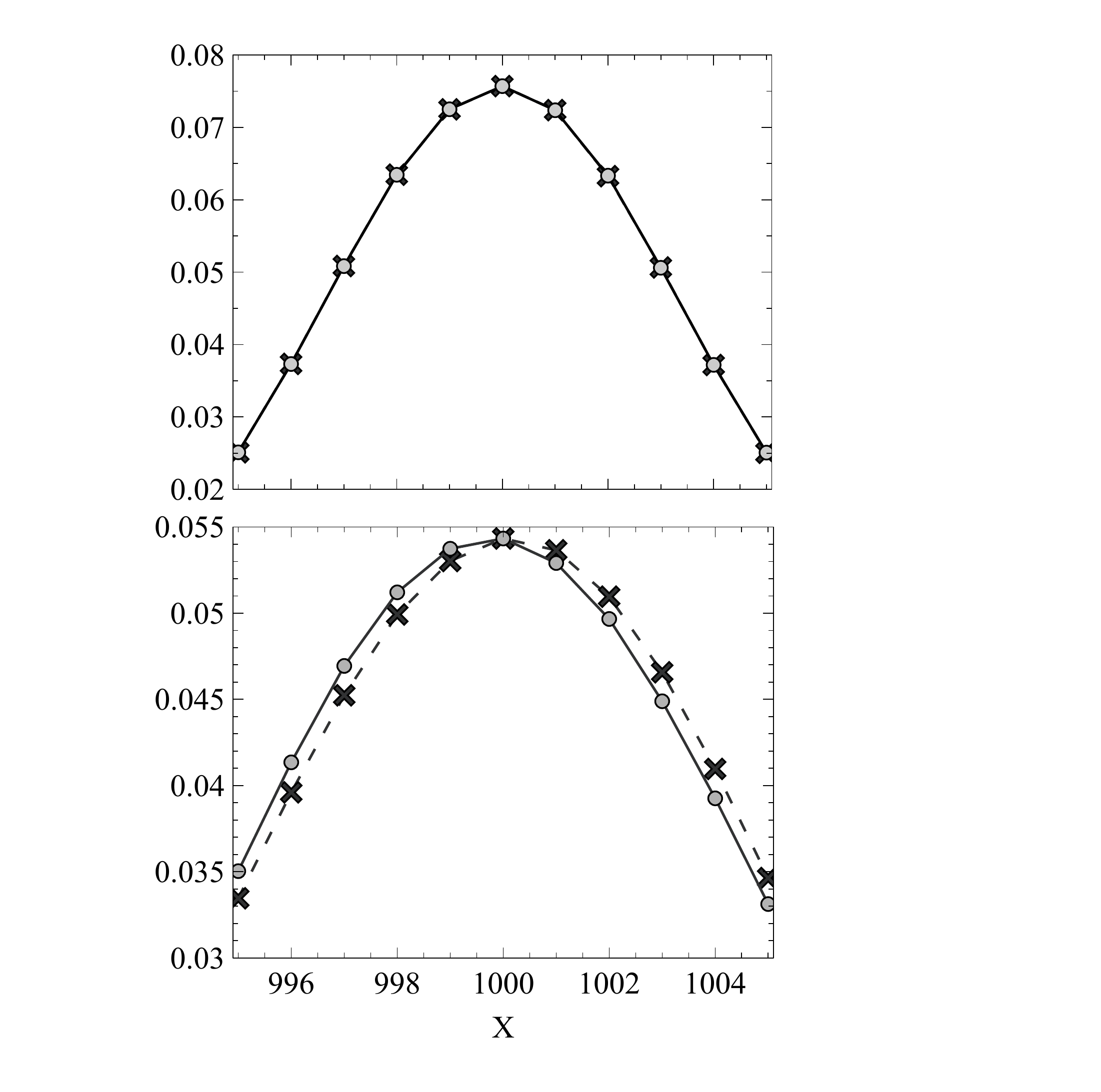,width=15.cm}
\caption{\label{fig:Probs}\small{Joint probability distributions $P(X, t \,\, ; \, X_{ref}, t + \tau)$ (crosses) and $P(X_{ref}, t \,\, ; \, X, t + \tau)$ (bullets), as a function of $X$, with $X_{ref} = 1000$ (macroscopic stationary state) and  $\tau = 10$  MRCT.  Top: model (\ref{Bd1}).  Bottom: model (\ref{Bd3}). Note that in both cases the entropy production $\sigma_s > 0$.}}
\label{figOne}
\end{center}
\end{figure}

\clearpage

\begin{figure}[h!]
\begin{center}
\epsfclipon
\epsfig{file=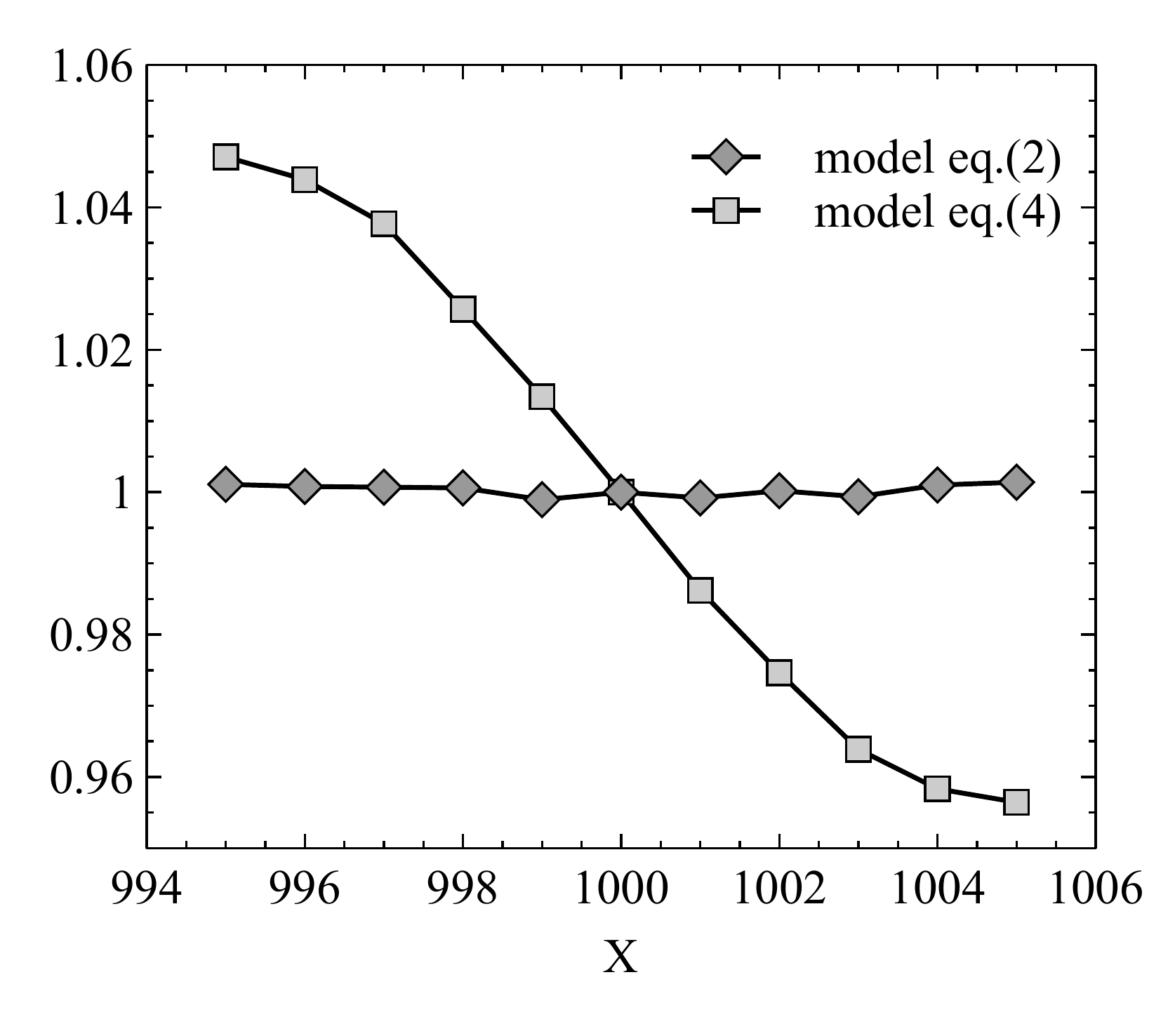,width=10.cm}
\caption{\label{fig:ProbRatio}\small{Probability ratio $P(X, t \,\, ; \, X_{ref}, t + \tau) \, / \, P(X_{ref}, t \,\, ; \, X, t + \tau)$ {(see the caption of Fig.~1)}  }}
\label{figTwo}
\end{center}
\end{figure}

\section{Concluding remarks and perspectives}

The study of the statistical properties of reactive systems is traditionally based on jump Markov process type of modeling, introduced in the mid-sixties \cite{McQuarrie:1967}.  As shown in \cite{Malek:2020,Malek:2017}, this traditional stochastic modeling limits the domain of applicability of the associated path thermodynamics to reactive systems involving only a single elementary reaction at the origin of each observed change in composition.  An alternative modeling that straightened this limitation was proposed recently \cite{Gaspard:2021}.  Using a  microscopic simulation of dilute reactive systems, in the Boltzmann limit, we showed in the present work that results obtained through this procedure are in excellent agreement with the predictions of the traditional modeling.

A peculiar consequence of this result concerns the class of one variable reactive systems in which all  reactions lead either to the change in composition $X \rightarrow X+1$ (forward) or $X \rightarrow X-1$ (backward).  A well-known example is the Schl\"ogl model \cite{Schlogl:1971}.  The model  (\ref{Bd1}) considered in our microscopic simulations is precisely of this type.
With the chosen parameter values for this model, the entropy production is strictly positive, which guarantees that the system operates under non-equilibrium conditions (cf. eq. (\ref{Bd2})).  However, the associated observed state trajectory proves to be time-reversible, in the sense that a sample path joining an arbitrary state $\Gamma_{1}$ to another arbitrary state $\Gamma_{\! 2}$ will occur with the same probability as the corresponding reverse path joining $\Gamma_{\! 2}$ to $\Gamma_1$.  But such a time-reversal symmetry is the key signature of thermodynamic equilibrium state where, on average, each forward reaction is exactly balanced by its reverse.  That is not the case here.   In addition, as shown previously \cite{Malek:2017,Malek:2020}, upon restricting ourselves to a well stirred system (no diffusion) and adopting a traditional Markovian modeling, the results observed in the simulation can be proved rigorously (see Section 6.3 in \cite{Gardiner:2009} for more details).  We are thus faced with a strange paradox.

Meticulous readers may object to our conclusions by pointing out that they may just result from an over simplified theoretical modeling of the system.  And they are right.  In particular,  a problematic feature concerns the "perfect homogeneity" assumption. Keeping a reactive system out of equilibrium requires fixing the concentration of some chemically active components to prescribed values.   Not only do these chemical intermediates fluctuate locally because of reactive collisions, even in well stirred systems, but they also diffuse.  Even though we are only interested in global (space averaged) quantities, there is {\it a priori} no guaranty that the effect of these local fluctuations cancels out through space averaging. In other words, local fluctuations could compromise the "perfect homogeneity" assumption.

Here again, in the absence of sufficiently accurate experimental results, the only way to address this issue is through microscopic simulations of reactive fluids.  For evident efficiency reasons, in the present article we used Bird's algorithm for microscopic simulations.  Not only this algorithm is up to 3 orders of magnitude faster than the traditional hard spheres molecular dynamics, but in addition it allows the simulation of perfectly homogeneous (zero dimensional) Boltzmann equation.  This choice was motivated by our main purpose:   investigate the validity of different types of stochastic modeling of a perfectly homogeneous reactive systems.  The next step is to perform microscopic simulations of reactive fluids using the exact hard spheres molecular dynamics procedure.  Work in this direction is in progress.

Finally, while numerical simulations are useful one should not conflate a computational algorithm with a mathematical process. For example, it is misleading (and incorrect) to define a jump Markov process as being equivalent to the Gillespie algorithm \cite{Gillespie:1976, Gillespie:1992} (see eqs. (2) and (3) in \cite{Gaspard:2021}).  Such a process is, in fact, defined by the "transition rate" concept, introduced by Kolmogorov \cite{Oksendal:2003}, while the algorithm is simply a numerical procedure to generate the so-called "minimal process" associated with jump Markov processes \cite{Karlin:1975}.

\section*{Acknowledgments}

The use of computational facilities at the Computing Center of the University of Bourgogne, DNUM-CCUB, is gratefully acknowledged. The authors thank Carmela Chateau-Smith for the careful reading of the manuscript. One author (AG) acknowledges support by the U.S. Department of Energy, Office of Science, Office of Advanced Scientific Computing Research, Applied Mathematics Program under contract No. DE-AC02-05CH11231.

\newpage

\section*{Appendix A: Direct Simulation Monte Carlo for Reactive Systems}

As in molecular dynamic simulation (MD), the state of the system in Direct Simulation Monte Carlo (DSMC) is the set of particle positions and velocities, $\{ {\bf r}_i, {\bf v}_i\}$. The evolution equations are integrated over successive time steps $\Delta t$, typically a fraction of the mean collision time for a particle. Within a time step, the free flight motion and the particle interactions (collisions) are assumed to be decoupled. The free flight motion for a particle $i$ is trivially computed as ${\bf r}_i(t + \Delta t) ={\bf r}_i(t) + {\bf v}_i(t) \Delta t$.  After all particles have been moved, they are sorted into "collisional" cells, typically a fraction of mean free path in length.  The main hypothesis in Bird's DSMC algorithm is that the cells are assumed to be perfectly homogeneous, i.e., all particles within a cell are considered to be potential collision partners, regardless of their exact positions. This basic hypothesis simplifies considerably the dynamics and allows the algorithm to be up to three orders of magnitude faster than the corresponding exact hard sphere MD.

A set of representative collisions, for the time step $\Delta t$, are then chosen in each cell. A collision probability is assigned to each selected pair based on their relative speed; a random impact parameter is selected and the collision is performed.  After the collision process has been completed in each cell, the particles are moved according to their updated velocities and the procedure is repeated.  At this point we may recall that our main purpose here is the study of the statistical properties of {\it perfectly homogeneous reactive systems}. The DSMC algorithm is particularly well adapted for this case since it allows the simulation of a homogeneous Boltzmann gas simply by associating the entire system volume to a single collisional cell.

Reactions are modeled using "hard sphere chemistry" which was introduced in the mid-1970's \cite{Portnow:1975,Boissonade:1979}.  The basic idea is quite elegant and simple. We first assign to each species an attribute, say a "color". A reactive collision occurs if the colliding particles have "enough" energy, i.e., if the relative kinetic energy of the colliding particles exceeds some threshold related to the activation energy of the reaction \cite{Present:1959}. If this is the case, then the colors of the particles are changed, according to the chemical step under consideration.

A major problem with this procedure is that it leads to the deformation of the Maxwell-Boltzmann distribution since only the most energetic particles can actually undergo a reactive transformation \cite{Prigogine:1949,Baras:1989}. To avoid this non-equilibrium effect, the frequency of reactive collisions must be significantly smaller than the frequency of elastic collisions, which results in an significant waste of CPU time. One way to overcome this difficulty is to further simplify the reactive collision rules by the following procedure. If the intensive quantities are expressed through mole fractions, instead of concentrations, then the kinetic constants are proportional to the collision frequency, that is
\begin{equation}
\label{ApA1}
k_i \, = \, \nu_i \, \exp{ \big\{- \, E_i / k_B \, T \, \big\} } \, \equiv \, \nu_i \, \widetilde{k}_i
\end{equation}
where $E_i$ is the activation energy of the reaction $i$ and $\nu_i$  is the collision frequency between the corresponding reactant particles.    After a collision between two such reactive particles has occurred, we choose randomly $\widetilde{k}_i \%$ of the collisions to be reactive, where $\widetilde{k}_i$ stands for the Arrhenius factor defined in eq. (\ref{ApA1}). Note that since the first reverse reaction in (\ref{Bd1}) involves a pair of the same particles (i.e.,  $X$ particles), the relation (\ref{ApA1}) must be replaced by $k_{- 1} =  \nu_{- 1} \, \widetilde{k}_{- 1} / 2$ for that reaction.  In any case,  this procedure avoids the deformation of the Maxwell-Boltzmann distribution since it does not involve any systematic energy transfer between reactants and products. It is, however, restricted to isothermal second-order (binary collisions) reactions (see \cite{Baras:1997} for a review).

A final issue concerns the appropriate microscopic procedure to maintain constant the mole fractions of some of the chemical species. To this end, in addition to the reactive chemicals, we also consider "solvent" particles $S$. Their role is precisely to maintain constant the number of $A$, $B$ and $C$ particles through the following strategy.  Each time one of these particles is created through a reactive collision, it is replaced by an $S$ particle.  Similarly,  when one of these particles is destroyed through a reactive collision, an $S$ particle is chosen randomly and transformed into that species.  Since  the solvent particles don't intervene directly in the reaction scheme (\ref{Bd1}), they don't modify the system's dynamics while maintaining it out of equilibrium \cite{Baras:1990}.  The very same way as the presence of particle reservoirs do, for example in CSTR (well stirred tank reactor). Finally, note that for both reaction models (\ref{Bd1}) and (\ref{Bd3}) the number of $A$, $B$, and $C$ particles and the sum of $X$ and solvent particles $X(t) + S(t)$ remain constant. As such, knowledge of $X(t)$  determines entirely the state of the system at each instant of time.


\newpage

\begin{thebibliography}{99}


\bibitem{Kondepudi:2008} D. Kondepudi, {\it Introduction to Modern Thermodynamics}, Wiley (2008).

\bibitem{Garcia:2022} A.L. Garcia, {\it Essentials of Modern Thermodynamics}, Amazon (2022).

\bibitem{Malek:2020} M. Malek Mansour and A. L. Garcia, Phys. Rev E, {\bf 101}, 052135 (2020).

\bibitem{Malek:2017} M. Malek Mansour and F. Baras, Chaos {\bf 27}, 104609 (2017).

\bibitem{DeDecker:2015} Y. De Decker, A. Garcia Cantu Ros and G. Nicolis, Euro. Phys. J. {\bf 224}, 947 (2015) ; Y. De Decker, J-F. Derivaux and G. Nicolis, Phys. Rev. E {\bf 93},  042127 (2016).

\bibitem{Seleznev:2015} D. Seleznev, G. A. Zhernokleeva and L. M. Martyushe, JETP Letters {\bf 102}, 557 (2015).

\bibitem{Kurchan:1998} J. Kurchan, J. Phys. A {\bf 31}, 3719 (1998) ; J. Stat. Mech., {\bf 07}, P07005 (2007).

\bibitem{Gaspard:2021} P. Gaspard, Phys. Rev E {\bf 103}, 016101 (2021).

\bibitem{Andrieux:2007}	 D. Andrieux and P. Gaspard, J. Stat. Phys. {\bf 127}, 107 (2007).

\bibitem{Lebowitz:1999}	 J. L. Lebowitz and H. Spohn, J. Stat. Phys., {\bf  95}, 333 (1999).

\bibitem{Gaspard:2004} P. Gaspard, J. Chem. Phys. {\bf 120}, 8898 (2004).

\bibitem{Andrieux:2004}	 D. Andrieux and P. Gaspard, J. Chem. Phys. {\bf 121}, 6167 (2004).

\bibitem{Oksendal:2003} See for example, B. {\O}ksendal, {\it Stochastic Differential Equations: An Introduction with Applications}. Springer (2003).

\bibitem{Ethier:2009}	S. N. Ethier and T. G. Kurtz, {\it Marko Processes: characterization and Convergence}, Wiley- Interscience (2009).

\bibitem{VanKampen:1983}	N. G. Van Kampen, {\it Stochastic Processes in Physics and
Chemistry}, North-Holland, Amsterdam (1983).

\bibitem{Gardiner:2009}	 C.W. Gardiner, {\it Handbook of Stochastic Methods}, Springer-Verlag (2009).

\bibitem{McQuarrie:1967} D. McQuarrie, Suppl. Rev. Ser: Appl. Prob., Methuen, London (1967).

\bibitem{Haken:1983} H. Haken, {\it Synergetics: An Introduction}, Springer-Verlag, Berlin, (1983).

\bibitem{Nicolis:1977} G. Nicolis and I. Prigogine, {\it Self-Organization in Nonequilibrium Systems}, Wiley- Interscience (1977).

\bibitem{Kurtz:1976}	T. G. Kurtz, Math. Progr. Stud. 5, {\bf 67} (1976) ;  Stoch. Proc. Appl. 6, {\bf 223} (1978) ; M. Malek Mansour, C. Van Den Broeck, G. Nicolis and J.W. Turner, Ann. Phys. (USA)  {\bf 131}, 283 (1981).

\bibitem{Seifert:2005} U. Seifert, Phys. Rev. Lett. {\bf 95}, 040602 (2005) ; Eur. Phys. J. B {\bf 64}, 423 (2008).

\bibitem{Portnow:1975}  J. Portnow, Phys. Lett. A, 51, 370 (1975).

\bibitem{Ortoleva:1976}  P. Ortoleva and S. Yip, J. Chem. Phys., {\bf 65}, 2045 (1976).

\bibitem{Boissonade:1979} J. Boissonade, Phys. Lett. {\bf A 74}, 285 (1979) ; Physica {\bf A 113}, 607 (1982).

\bibitem{Baras:1989} F. Baras and M. Malek Mansour, Phys. Rev. Lett., {\bf 63}, 2429 (1989); M. Malek Mansour and F. Baras, Physica A, {\bf 188}, 253 (1992).

\bibitem{Lemarchand:1999} A. Lemarchand and B. Nowakowski, J. Chem. Phys. {\bf 111}, 6190 (1999) ; EuroPhys. Lett. {\bf 94}, 48004 (2011).

\bibitem{Hansen:2006}  J. S. Hansen, B. Nowakowski and A. Lemarchand, J. Chem. Phys. {\bf 124}, 034503 (2006) ; P. Dziekan, L. Signon, B. Nowakowski, and A. Lemarchand, J. Chem. Phys. {\bf 139}, 114107 (2013).

\bibitem{Baras:2004} F. Baras, M. Salazar, E. Kestemont and M. Malek Mansour, Europhys. Lett., {\bf 67} (6), 900-906 (2004).

\bibitem{Baras:1997} F. Baras and M. Malek Mansour, Adv. Chem. Phys. {\bf 100}, 393 (1997).

\bibitem{Schlogl:1971} F. Schl\"{o}gl, Z.	Phys. {\bf 248}, 446 (1971) ; Z.	Phys. {\bf 253}, 147 (1972).

\bibitem{Graham:1971} R. Graham and H. Haken, {\it Generalized thermodynamic potential for Markov system in detailed balance and far from thermal equilibrium}, Z. Phys. {\bf 243} 289 (1971).

\bibitem{Bird:1976} G.A. Bird, {\it Molecular Gas Dynamics} (Clarendon, Oxtord, 1976).

\bibitem{Garcia:2000} A.L. Garcia, {\it Numerical methods for physics}, Englewood Cliffs, NJ: Prentice Hall, (2000).

\bibitem{Gillespie:1976}   D. T. Gillespie, J. Comput. Phys. {\bf 22}, 403 (1976) ; J. Phys. Chem. {\bf 81}, 2340
(1977).

\bibitem{Gillespie:1992}   D. T. Gillespie, {\it Markov Processes: An Introduction of Physical Scientists} (Academic, New York, 1992).

\bibitem{Karlin:1975}  S. Karlin and H. Taylor, {\it A First Course in Stochastic Processes}, (Academic Press, 1975).

\bibitem{Present:1959} R. D. Present, J. Chem. Phys., {\bf 31}, 747 (1959).

\bibitem{Prigogine:1949} I. Prigogine and E. Xhrouet, Physica, {\bf 15}, 913 (1949).

\bibitem{Baras:1990} F. Baras, J. E. Pearson, and M. Malek Mansour, J. Chem. Phys., {\bf 93}, 5747 (1990).


\end{thebibliography}
\end{document}